\documentclass[journal=jctc,manuscript=article]{achemso}
\usepackage[utf8]{inputenc}
\usepackage[T1]{fontenc}
\usepackage{xcolor}
\usepackage{amsmath}
\usepackage{amssymb}
\usepackage{graphicx}
\usepackage{multirow}
\usepackage[version=4]{mhchem}
\usepackage[section]{placeins}
\usepackage{float}
\usepackage{booktabs}
\usepackage{silence}
\usepackage{array}
\usepackage{braket}
\newcolumntype{K}{>{\centering\arraybackslash}b{0.85cm}}
\newcolumntype{L}{>{\centering\arraybackslash}b{1.28cm}}
\usepackage[english]{babel}
\usepackage{subfigure}
\usepackage{microtype}
\usepackage{longtable}
\SectionNumbersOn


\title{Combining Renormalized Singles $GW$ Methods with the Bethe-Salpeter Equation for Accurate Neutral Excitation Energies}

\author{Jiachen Li}
\affiliation{Department of Chemistry, Duke University, Durham, NC 27708, USA}
\author{Dorothea Golze}
\affiliation{Faculty of Chemistry and Food Chemistry, Technische Universit\"at Dresden, 01062 Dresden, Germany}
\author{Weitao Yang}
\affiliation{Department of Chemistry, Duke University, Durham, NC 27708, USA}
\email{weitao.yang@duke.edu}

\begin{document}

\begin{tocentry}
\includegraphics[width=1\textwidth]{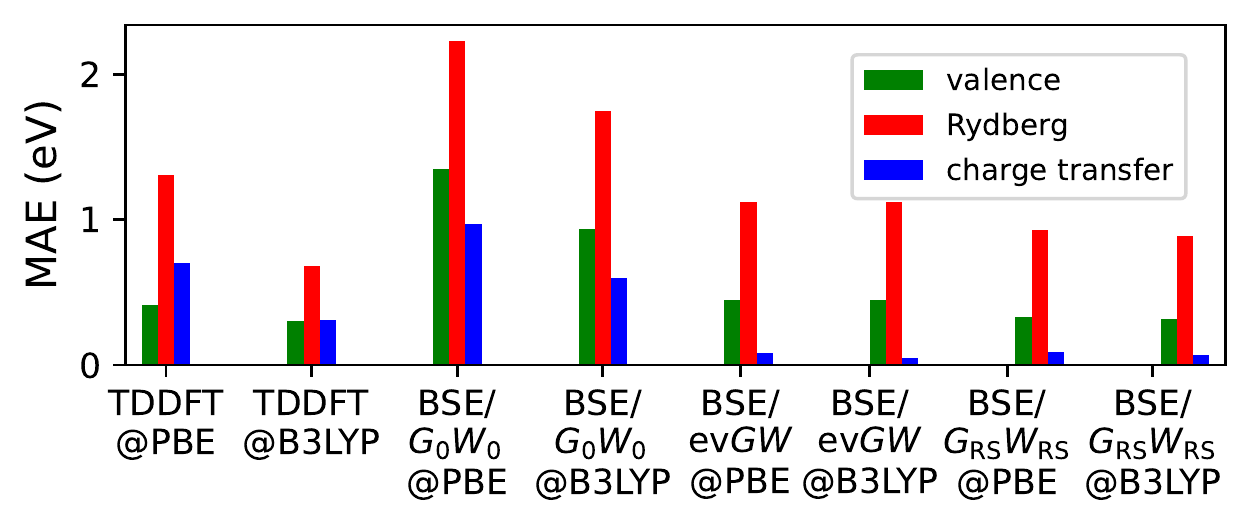}
\end{tocentry}

\begin{abstract}
We apply the renormalized singles (RS) Green's function in the Bethe-Salpeter equation (BSE)/$GW$ approach to predict accurate neutral excitation energies of molecular systems.
The BSE calculations are performed on top of the $G_{\text{RS}}W_{\text{RS}}$ method, which uses the RS Green's function also for the computation of the screened Coulomb interaction $W$.
We show that the BSE/$G_{\text{RS}}W_{\text{RS}}$ approach significantly outperforms BSE/$G_0W_0$ for predicting excitation energies of valence, Rydberg and charge transfer (CT) excitations by benchmarking the Truhlar-Gagliardi set, Stein CT set and an atomic Rydberg test set.
For the Truhlar-Gagliardi test set,
BSE/$G_{\text{RS}}W_{\text{RS}}$ provides comparable accuracy to time-dependent density functional theory (TDDFT) and is slightly better than BSE starting from eigenvalue self-consistent $GW$ (ev$GW$).
For the Stein CT test set,
BSE/$G_{\text{RS}}W_{\text{RS}}$ significantly outperforms BSE/$G_0W_0$ and TDDFT with the accuracy comparable to BSE/ev$GW$.
We also show that BSE/$G_{\text{RS}}W_{\text{RS}}$ predicts Rydberg excitation energies of atomic systems well.
Besides the excellent accuracy,
BSE/$G_{\text{RS}}W_{\text{RS}}$ largely eliminates the dependence on the choice of the density functional approximation.
This work demonstrates that the BSE/$G_{\text{RS}}W_{\text{RS}}$ approach is accurate and efficient for predicting excitation energies for a broad range of systems,
which expands the applicability of the BSE/$GW$ approach.
\end{abstract}

\maketitle

\section{INTRODUCTION}
The accurate prediction of neutral excitation energies from first-principle calculations plays a critical role in guiding understanding and new developments in chemistry, biochemistry and material science,
such as drug delivery\cite{velemaOpticalControlAntibacterial2013,beharryAzobenzenePhotoswitchingUltraviolet2011}
and design of organic sensitizer\cite{shalabiEffectsMacrocycleAnchoring2015,zhangDFTTDDFTStudy2010}.
In the past decades,
many theoretical approaches have been developed to compute accurate excitation energies. Time-dependent density functional theory\cite{rungeDensityFunctionalTheoryTimeDependent1984, casidaTimeDependentDensityFunctional1995, ullrichTimeDependentDensityFunctionalTheory2011} (TDDFT) ranks among the most popular approaches.
Because of the good accuracy and the affordable computational cost,
TDDFT has been widely implemented in modern quantum chemistry packages to calculate energies, structures and other properties of excited states for molecular and periodic systems\cite{casidaTimedependentDensityfunctionalTheory2009,casidaLinearResponseTimeDependentDensity2006,laurentTDDFTBenchmarksReview2013}.
However, TDDFT has several well-known shortcomings.
For example,
TDDFT with commonly used density functional approximations (DFAs) poorly predicts excitation energies of Rydberg and charge transfer (CT) excitations\cite{tozerRelationshipLongrangeChargetransfer2003, dreuwLongrangeChargetransferExcited2003}.
The failure of TDDFT for these excitations must be attributed to the incorrect description of the long-range behavior using conventional DFAs\cite{dreuwLongrangeChargetransferExcited2003,tozerRelationshipLongrangeChargetransfer2003}. Attempts to address this issue include
the usage of range-separated or Coulomb-attenuated functionals\cite{leiningerCombiningLongrangeConfiguration1997,besleyTimedependentDensityFunctional2009,peachExcitationEnergiesDensity2008,aiRoleRangeSeparatedCorrelation2021}
and tuning the fraction of the Hartree-Fock (HF)\cite{slaterNoteHartreeMethod1930,szaboModernQuantumChemistry2012} exchange in DFAs\cite{zhaoDensityFunctionalSpectroscopy2006}.
In addition,
the accuracy of TDDFT strongly depends on the exchange-correlation (XC) kernel,
which is the second derivative of the XC energy from the given DFA with respect to the electron density.
\cite{laurentTDDFTBenchmarksReview2013, leangBenchmarkingPerformanceTimedependent2012}
The difference of calculated excitation energies obtained from TDDFT with different DFAs can exceed $1.0$ \,{eV} for valence excitations and even exceed $2.0$ \,{eV} for Rydberg excitations\cite{kaurWhatAccuracyLimit2019,laurentTDDFTBenchmarksReview2013}.
Recent developments to improve the accuracy of TDDFT include multireference DFT\cite{chenMultireferenceDensityFunctional2017,liMultireferenceDensityFunctional2022},
orbital optimized (OO) DFT\cite{haitOrbitalOptimizedDensity2021}
and mixed-reference spin-flip (MRSF)-TDDFT\cite{horbatenkoMixedReferenceSpinFlipTimeDependent2021}. \\

In the last years, the Bethe-Salpeter equation\cite{shamManyParticleDerivationEffectiveMass1966, hankeManyParticleEffectsOptical1979, salpeterRelativisticEquationBoundState1951} (BSE) formalism in the Green's function many-body perturbation theory\cite{hedinNewMethodCalculating1965, martinInteractingElectrons2016} (MBPT) has become increasingly popular for computing the optical excitations of molecules\cite{blaseBetheSalpeterEquation2020} and has been most recently also applied to $K$-edge transitions.\cite{yaoAllElectronBSEGW2022}
BSE is commonly performed on top of a $GW$ calculation,\cite{martinInteractingElectrons2016, reiningGWApproximationContent2018, golzeGWCompendiumPractical2019} using the $GW$ quasiparticle (QP) energies as input. We denote this approach as BSE/$GW$.
Recently, BSE combining with the T-matrix approximation for predicting neutral excitation energies has also been reported\cite{loosStaticDynamicBethe2022}.
In the BSE/$GW$ approach, the electron-hole interaction is described by the screened Coulomb interaction $W$ instead of the bare Coulomb interaction.
The $GW$ QP energies substantially improve upon the Kohn-Sham (KS) orbital energies for predicting fundamental HOMO-LUMO gaps and ionization potentials in general,
which are the critical quantities in BSE to calculate excitation energies.
BSE/$GW$ has been successfully applied to calculate excitation energies for systems of different sizes including molecules,
solids and low-dimensional materials\cite{jacqueminBenchmarkBetheSalpeterTriplet2017, escuderoModelingPhotochromeTiO22017, jacqueminBetheSalpeterFormalism2017, azariasBetheSalpeterStudyCationic2017, azariasCalculationsTransitionEnergies2017, blaseChargetransferExcitationsMolecular2011, gattiElectronholeInteractionsCorrelated2015, cudazzoExcitonBandStructure2016, sottileEfficientInitioCalculations2007, jiangRealtimeGWBSEInvestigations2021, liuAllelectronInitioBetheSalpeter2020,yaoAllElectronBSEGW2022,dvorakQuantumEmbeddingTheory2019,moninoSpinConservedSpinFlipOptical2021,loosDynamicalCorrectionBethe2020}.\\

However, the BSE/$GW$ approach still suffers from several problems.
First,
although BSE has a computationally favorable scaling of $\mathcal{O}(N^4)$ with respect to the system size $N$,
which is the same as in TDDFT,
the preceding $GW$ calculation is computationally demanding.
In the fully analytical treatment of the $GW$ self-energy,
the diagonalization of the random phase approximation (RPA) equation scales as $\mathcal{O}(N^6)$ and the self-energy evaluation scales as $\mathcal{O}(N^5)$\cite{golzeGWCompendiumPractical2019,vansettenGWMethodQuantumChemistry2013}.
Thus, the computationally expensive $GW$ step dominates the computational cost in a BSE/$GW$ calculation.
Many techniques have been developed to reduce the computational cost of $GW$ calculations, e.g., the contour deformation technique\cite{golzeCoreLevelBindingEnergies2018} or analytic continuation of the self-energy\cite{renResolutionofidentityApproachHartree2012,ducheminRobustAnalyticContinuationApproach2020}, which scale $\mathcal{O}(N^4)$ for valence states.
The computational cost is further reduced in cubic scaling $GW$ implementations, which recently emerged for localized basis set codes\cite{wilhelmGWCalculationsThousands2018,Foerster2020,wilhelmLowScalingGWBenchmark2021,ducheminCubicScalingAllElectronGW2021} enabling
$GW$ calculations for large systems of more than thousand atoms.

Second,
the accuracy of the commonly used BSE/$G_0W_0$ approach has an undesirable dependence on the choice of the DFA\cite{ziaeiGWBSEApproachS12016,liCombiningLocalizedOrbital2022},
which is due the perturbative nature of the $G_0W_0$ scheme.
In BSE/$G_0W_0$, the dependence on the DFA starting point is in the range of $0.5$ \,{eV} for predicting valence excitation energies and can even exceed $1.0$ \,{eV} for predicting CT excitation energies\cite{liCombiningLocalizedOrbital2022}.
It has been shown that BSE/$G_0W_0$ based on range-separated functionals and tuned hybrid functionals provides more accurate excitation energies\cite{brunevalSystematicBenchmarkInitio2015,yaoAllElectronBSEGW2022} than BSE/$G_0W_0$ based on GGA functionals.
This dependence can be largely reduced by introducing self-consistency into the $GW$ calculations,
such as the eigenvalue-self-consistent $GW$ (ev$GW$) method, where the eigenvalues are iterated in $G$ and $W$, the quasiparticle-self-consistent $GW$ (qs$GW$) scheme\cite{vanschilfgaardeQuasiparticleSelfConsistentGW2006,kaplanQuasiParticleSelfConsistentGW2016}  or the fully self-consistent $GW$ (sc$GW$) approach\cite{carusoUnifiedDescriptionGround2012, carusoSelfconsistentGWAllelectron2013}.
It has been shown that the BSE/ev$GW$ approach provides an accuracy comparable to TDDFT for predicting valence excitations and significantly outperforms BSE/$G_0W_0$ and TDDFT for predicting CT and Rydberg excitations\cite{jacqueminAssessmentConvergencePartially2016,jacqueminBetheSalpeterFormalism2017,liCombiningLocalizedOrbital2022}.
The DFA dependence in BSE/ev$GW$ is largely washed out\cite{jacqueminAssessmentConvergencePartially2016,jacqueminBetheSalpeterFormalism2017,liCombiningLocalizedOrbital2022}.
In practice, ev$GW$ calculations can converge within a few steps by using the linear mixing method\cite{kaplanQuasiParticleSelfConsistentGW2016}.
However, ev$GW$ might have convergence difficulties for systems with multiple solutions\cite{verilUnphysicalDiscontinuitiesGW2018,moninoUnphysicalDiscontinuitiesIntruder2022}.
In addition, the extra computational cost in ev$GW$ is expensive for large systems.

To reduce the computational cost, efforts were recently made to approximate the $GW$ QP energies by improved orbital energies from DFT and to use the latter as BSE input.
For example, we recently employed the localized orbital scaling correction\cite{liLocalizedOrbitalScaling2018} (LOSC), which minimizes the delocalization error in DFAs to predict orbital energies. We showed that the BSE/LOSC approach provides good accuracy for predicting excitation energies of different systems\cite{liCombiningLocalizedOrbital2022} and yields significantly better results than BSE/$G_0W_0$. The BSE/LOSC approach scales as $\mathcal{O}(N^4)$ complexity.
Another computationally cheaper BSE approach is based on the Koopmans-compliant (KC) functionals, where the orbital energies are derived from KC functionals and the screened interaction is obtained via a direct minimization on top of a maximally localized Wannier function basis\cite{elliottKoopmansMeetsBethe2019}. The KC-based BSE yields similar a accuracy as the BSE/$G_0W_0$ method\cite{elliottKoopmansMeetsBethe2019}. \\

In this work,
we applied the recently developed renormalized singles (RS) Green's function\cite{jinRenormalizedSinglesGreen2019,li2022benchmark} in BSE/$GW$ to compute accurate excitation energies with affordable computational cost.
The idea of the RS Green's function approach is to compute the HF self-energy with the KS orbitals from DFT\cite{hohenbergInhomogeneousElectronGas1964, kohnSelfConsistentEquationsIncluding1965, parrDensityFunctionalTheoryAtoms1989} instead of the HF orbitals.
Because of Brillouin's theorem\cite{szaboModernQuantumChemistry2012},
there is no singles contribution to the HF self-energy in the perturbation.
However, the HF Green's function is not a good starting point for $G_0W_0$ calculations\cite{brunevalBenchmarkingStartingPoints2013, jinRenormalizedSinglesGreen2019}.
Therefore, the HF Hamiltonian is constructed with KS orbitals and diagonalized separately in the occupied and virtual subspaces.
This renormalization procedure absorbs all singles contributions into the self-energy to reduce the dependence on the choice of the DFA.
The resulting RS Green's function is constructed with RS orbital energies from the renormalization process,
which has the same form as the KS Green's function.
From the viewpoint of the RPA,
higher order contributions are also included in the infinite summation by this renormalization process.
Compared with the self-consistent $GW$ methods that solve Hedin's equations iteratively to eliminate the starting point dependence,
the one-shot RS process captures all the contributions of the single excitations while hardly increasing the computational compared to $G_0W_0$.
The RS Green's function has been used in the $GW$ methods and the T-matrix methods to predict accurate quasiparticle energies for valence and core states\cite{jinRenormalizedSinglesGreen2019,liRenormalizedSinglesGreen2021,liCombiningLocalizedOrbital2022}.
The concept of RS has also been used in the multireference DFT approach\cite{liMultireferenceDensityFunctional2022} to describe the static correlation in strongly correlated systems.
The RS Green's function shares similar thinking as the renormalized single-excitation (rSE) in the RPA calculation for correlation energies\cite{renRandomPhaseApproximationElectron2011, renRenormalizedSecondorderPerturbation2013, paierAssessmentCorrelationEnergies2012}.  \\

The RS Green's function has been applied in the $GW$ approximation in two flavors.
The first one is the $G_{\text{RS}}W_0$\cite{jinRenormalizedSinglesGreen2019} method, which uses the RS Green's function as a new starting point and calculates the screened interaction with the KS Green's function.
We found that the $G_{\text{RS}}W_0$ method provides a considerable improvement over $G_0W_0$ for predicting valence ionization potentials and electron affinities with a reduced starting point dependence\cite{jinRenormalizedSinglesGreen2019}, but fails to restore the correct physics for deep core excitations\cite{li2022benchmark}.
The second one is the $G_{\text{RS}}W_{\text{RS}}$ method\cite{li2022benchmark}, which uses the RS Green's function as a new starting point and calculates the screened interaction with the RS Green's function. We found that $G_{\text{RS}}W_{\text{RS}}$ also yields an improvement over $G_0W_0$\cite{li2022renormalized}, i.e., it opens the fundamental gaps compared to $G_0W_0$ similarly as ev$GW$, see also Section~1 in the Supporting information.
Because of the non-linear nature of the QP equation, multiple solutions can be found due to the unphysical discontinuities and intruder states\cite{loosGreenFunctionsSelfConsistency2018,moninoUnphysicalDiscontinuitiesIntruder2022,verilUnphysicalDiscontinuitiesGW2018}, especially for core states. $G_{\text{RS}}W_{\text{RS}}$ properly separates the main core QP state from the satellites and provides a solution to the QP equation corresponding to the desired core state\cite{li2022benchmark}, which is not the case for $G_0W_0$ with GGA or hybrid functionals with a low amount of exact exchange. Even though the absolute core-level energies are overestimated by several electronvolts, the relative shifts are predicted with a reasonable accuracy\cite{li2022benchmark}.
For valence as well as core states, the starting point dependence is significantly reduced compared to $G_0W_0$.\\

In the present work, we benchmark both RS flavors, BSE/$G_{\text{RS}}W_0$ and BSE/$G_{\text{RS}}W_{\text{RS}}$ for the prediction of neutral excitations of molecular systems, including valence, CT and Rydberg excitations.

\section{THEORY}

\subsection{RS Green's Function}\label{subsec:rs}

The goal of the RS Green's function approach is to improve the accuracy of perturbative $GW$ methods by including exactly the exchange self energy, which is an one-electron contribution, and to reduce the starting point dependence of the orbital energies on the DFA. The general idea is to construct
the HF self-energy from KS orbitals,
followed by a separate diagonalization in the  occupied orbital subspace and the virtual orbital subspace\cite{jinRenormalizedSinglesGreen2019}.
Effectively, the RS approach treats the one-body exchange self energy exactly,
or non-perturbatively,
by a diagonalization,
unlike the commonly used $G_0W_0$ approach,
in which the one-body exchange is included perturbatively along with the many-body correlation self energy.
The RS Green's function $G_{\text{RS}}$ is defined as the solution of the two projected equations in the occupied orbital subspace and the virtual orbital subspace\cite{jinRenormalizedSinglesGreen2019}
\begin{equation}
P(G_{\text{RS}}^{-1})P = P(G_{0}^{-1})P + P(\Sigma_{\text{Hx}}[G_{0}]-v_{\text{Hxc}})P \text{,}
\end{equation}
and
\begin{equation}
Q(G_{\text{RS}}^{-1})Q = Q(G_{0}^{-1})Q + Q(\Sigma_{\text{Hx}}[G_{0}]-v_{\text{Hxc}})Q \text{,}
\end{equation}
where $P=\sum_{i}^{\text{occ}}|\psi_{i}\rangle\langle\psi_{i}|$ is the projection into the occupied orbital space and $Q=I-P$ is the projection into the virtual orbital space, $\psi$ is the KS orbital, $\Sigma_{\text{Hx}}$ is the Hartree-exchange self-energy and $v_{\text{Hxc}}$ is the Hartree-exchange-correlation potential.
$\Sigma_{\text{Hx}}[G_{0}]$ means that the HF self-energy (Hartree and exchange contribution) is constructed from the KS density matrix.
Equivalently, the RS Green's function is obtained by using the DFA density matrix in the HF Hamiltonian,
namely $H_{\text{HF}}[G_{0}]$,
and solving two projected HF equations in the occupied and the virtual subspaces\cite{jinRenormalizedSinglesGreen2019}
\begin{equation}
    P(H_{\text{HF}}[G_{0}])P|\psi_{i}^{\text{RS}}\rangle =
    \epsilon_{i}^{\text{RS}}P|\psi_{i}^{\text{RS}}\rangle \text{,}\label{eq:rs_occ}
\end{equation}
and
\begin{equation}
    Q(H_{\text{HF}}[G_{0}])Q|\psi_{a}^{\text{RS}}\rangle =
    \epsilon_{a}^{\text{RS}}Q|\psi_{a}^{\text{RS}}\rangle \text{,}\label{eq:rs_vir}
\end{equation}
where $\epsilon^{\text{RS}}$ is the RS orbital energy and $\psi^{\text{RS}}$ is the RS orbital.
Here, we use $i$, $j$, $k$, $l$ for occupied orbitals, $a$, $b$, $c$, $d$ for virtual orbitals, $p$, $q$, $r$, $s$ for general orbitals.
The subspace diagonalization of the HF Hamiltonian is performed only once.
The resulting RS Green's function is diagonal in the occupied and virtual subspaces\cite{jinRenormalizedSinglesGreen2019}
\begin{equation}\label{eq:rs_green}
    [G_{\text{RS}}(\omega)]_{pq} = \delta_{pq}
    \frac{1}{\omega-\epsilon_{p}^{\text{RS}} +
    i\eta\text{sgn}(\epsilon_{p}^{\text{RS}}-\mu)}\text{.}
\end{equation}
Here $\mu$ is the chemical potential and $\eta$ is the broadening parameter.
As shown in Eq.\ref{eq:rs_green},
the RS Green's function has a similar form as the KS Green's function except that the KS orbital energies in the denominator are replaced by the RS orbital energies.
Therefore, the RS Green's function can be directly implemented in existing $GW$ codes.

\subsection{$G_{\text{RS}}W_0$ and $G_{\text{RS}}W_{\text{RS}}$}\label{subsec:rsgw}

The RS Green's function is used in two $GW$ flavors:
$G_{\text{RS}}W_0$\cite{jinRenormalizedSinglesGreen2019} and $G_{\text{RS}}W_{\text{RS}}$\cite{li2022benchmark}.
In both methods,
KS orbitals instead of RS orbitals are used for simplicity\cite{jinRenormalizedSinglesGreen2019,li2022benchmark} because using the RS orbitals does not change the results significantly.
Therefore,
the exchange part of the self-energy in $G_{\text{RS}}W_0$ and $G_{\text{RS}}W_{\text{RS}}$ is the same as $G_0W_0$.
In $G_{\text{RS}}W_0$\cite{jinRenormalizedSinglesGreen2019},
the RS Green's function is used as the starting point and the screened interaction is calculated with the KS Green's function,
which means the KS Green's function is used in the RPA calculation.
The correlation part of the self-energy in $G_{\text{RS}}W_0$ is\cite{jinRenormalizedSinglesGreen2019}
\begin{equation}\label{eq:grsw0_se}
        [\Sigma^{G_{\text{RS}}W_0}_{\text{c}}(\omega)]_{pp} =
        \sum_m \sum_q \frac{|(pq|\rho^{\text{KS}}_m)|^2}{\omega - \epsilon^{\text{RS}}_q
        - (\Omega^{\text{KS}}_m - i\tilde{\eta}) \text{sgn} (\epsilon^{\text{RS}}_q - \mu)} \text{, }
\end{equation}
where $\rho^{\text{KS}}_m$ and $\Omega^{\text{KS}}_m$ are the transition density and the excitation energy from RPA calculated with the KS Green's function,
$m$ is the index for the RPA excitation,
and $\tilde{\eta}=3\eta$\cite{vansettenGWMethodQuantumChemistry2013}. \\

With the self-energy in Eq.\ref{eq:grsw0_se},
the QP equation for $G_{\text{RS}}W_0$ is\cite{jinRenormalizedSinglesGreen2019}
\begin{equation}\label{eq:qpe_grsw0}
    \epsilon^{\text{QP}}_{p}=\epsilon_{p}^{0} + \langle p |
    \Sigma^{G_{\text{RS}}W_0}_{\text{xc}}(\epsilon_{p}^{\text{QP}})
    - v_{\text{xc}}|p\rangle \text{.}
\end{equation}
where $\epsilon^0_p$ is the KS orbital energy.
In Eq.\ref{eq:qpe_grsw0} the QP energy $\epsilon^{\text{QP}}_p$ appears in both sides,
which means Eq.\ref{eq:qpe_grsw0} needs to be solved iteratively.
To reduce the computational cost,
Eq.\ref{eq:qpe_grsw0} can be linearized\cite{martinInteractingElectrons2016, jinRenormalizedSinglesGreen2019} as
\begin{equation}
    \epsilon_{p}^{\text{QP}} = \epsilon_{p}^{\text{0}}
    + Z_{p}^{G_{\text{RS}}W_{0}}\langle p| \Sigma_{\text{xc}}^{G_{\text{RS}}W_{0}}(\epsilon_{p}^{\text{RS}}) - v_{\text{xc}}|p\rangle \text{,} \label{eq:qpe_grsw0_linear}
    \end{equation}
with the factor $Z_{p}^{G_{\text{RS}}W_{0}}=(1-\frac{\partial [\Sigma_{\text{c}}^{\text{\ensuremath{G_{\text{RS}}W_{0}}}}(\omega)]_{pp}}{\partial\omega}|_{\omega=\epsilon_{p}^{\text{RS}}})^{-1}$. \\

The $G_{\text{RS}}W_{\text{RS}}$ method\cite{li2022benchmark} uses the RS Green's function as a new starting point and calculates the screened interaction with the RS Green's function,
which means that the RS Green's function is used in the RPA calculation.
The correlation part of the self-energy in $G_{\text{RS}}W_{\text{RS}}$\cite{li2022benchmark}
\begin{equation}\label{eq:grswrs_se}
        [\Sigma^{G_{\text{RS}}W_{\text{RS}}}_{\text{c}}(\omega)]_{pp} =
        \sum_m \sum_q \frac{|(pq|\rho^{\text{RS}}_m)|^2}{\omega - \epsilon^{\text{RS}}_q
        - (\Omega^{\text{RS}}_m - i\tilde{\eta}) \text{sgn} (\epsilon^{\text{RS}}_q - \mu)} \text{, }
\end{equation}
where $\rho^{\text{RS}}_m$ and $\Omega^{\text{RS}}_m$ are the transition density
and the excitation energy from RPA calculated with the RS Green's function. \\

Therefore, the QP equation for $G_{\text{RS}}W_{\text{RS}}$ is\cite{li2022benchmark}
\begin{equation}\label{eq:qpe_grswrs}
    \epsilon^{\text{QP}}_{p}=\epsilon_{p}^{0} + \langle p |
    \Sigma_{\text{xc}}^{G_{\text{RS}}W_{\text{RS}}}(\epsilon_{p}^{\text{QP}}) -
    v_{\text{xc}}|p\rangle \text{.}
\end{equation}
Eq.\ref{eq:qpe_grswrs} can also be linearized as,
\begin{equation}
    \epsilon_{p}^{\text{QP}} = \epsilon_{p}^{\text{0}}
    + Z_{p}^{G_{\text{RS}}W_{\text{RS}}}\langle p|\Sigma_{\text{xc}}^{G_{\text{RS}}W_{\text{RS}}}(\epsilon_{p}^{\text{RS}})-v_{\text{xc}}|p\rangle \text{,} \label{eq:qpe_grswrs_linear}
    \end{equation}
with the factor $Z_{p}^{G_{\text{RS}}W_{\text{RS}}}=(1-\frac{\partial [\Sigma_{\text{c}}^{\text{\ensuremath{G_{\text{RS}}W_{\text{RS}}}}}(\omega)]_{pp}}{\partial\omega}|_{\omega=\epsilon_{p}^{\text{RS}}})^{-1}$. \\

As shown in Ref.\citenum{golzeGWCompendiumPractical2019},
the linearized QP equation gives small errors for valence QP energy calculations,
which are important in BSE.
In Section~2 in the Supporting Information,
we show that using linearized QP equations defined in Eq.\ref{eq:qpe_grsw0_linear} and Eq.\ref{eq:qpe_grswrs_linear} gives small differences around $0.01$ \,{eV} compared to Eq.\ref{eq:qpe_grsw0} and Eq.\ref{eq:qpe_grswrs} for the type of excitations studied here.
Therefore, the linearized QP equations are solved to reduce the computational cost in the present work.\\

\subsection{BSE/$G_{\text{RS}}W_0$ and BSE/$G_{\text{RS}}W_{\text{RS}}$ Approaches}\label{subsec:bse_rsgw}
The QP energies obtained from $G_{\text{RS}}W_{\text{RS}}$ are used in BSE to calculate excitation energies.
With the static approximation for the screened interaction that treats the frequency as zero\cite{krauseImplementationBetheSalpeter2017,blaseBetheSalpeterEquation2020,ghoshConceptsMethodsModern2016},
the working equation of BSE is a generalized eigenvalue equation\cite{krauseImplementationBetheSalpeter2017,ghoshConceptsMethodsModern2016,blaseBetheSalpeterEquation2020},
which is similar to the Casida equation in TDDFT\cite{ullrichTimeDependentDensityFunctionalTheory2011,casidaTimeDependentDensityFunctional1995}
\begin{equation}\label{eq:bse}
    \begin{bmatrix}
        \mathbf{A} & \mathbf{B} \\
        \mathbf{B^*} & \mathbf{A^*}
    \end{bmatrix}
    \begin{bmatrix}
        \mathbf{X} \\
        \mathbf{Y}
    \end{bmatrix}
    = \Omega
    \begin{bmatrix}
        \mathbf{I} & \mathbf{0} \\
        \mathbf{0} & \mathbf{-I}
    \end{bmatrix}
    \begin{bmatrix}
        \mathbf{X} \\
        \mathbf{Y}
    \end{bmatrix}\text{.}
\end{equation}
where $\Omega$ is the excitation energy.
In Eq.\ref{eq:bse} the $\mathbf{A}$, $\mathbf{B}$ matrices are defined as
\begin{align}
    A_{ia,jb} &= \delta_{ij} \delta_{ab} (\epsilon_a^{\text{QP}}-\epsilon_i^{\text{QP}}) + v_{ia,jb} - W_{ij,ab}(\omega=0) \text{,}\\
    B_{ia,jb} &= v_{ia,bj} - W_{ib,aj}(\omega=0) \text{,}
\end{align}
where $v$ is the Coulomb interaction and $W(\omega=0)$ is the static screened interaction.
$v$ is the Coulomb interaction defined as
\begin{equation}
    v_{pq,rs} =
    \int dx_{1}dx_{2}\frac{\psi_{p}^{*}(x_{1})\psi_{r}^{*}(x_{2})\psi_{q}(x_{1})\psi_{s}(x_{2})}{|\mathbf{r_{1}}-\mathbf{r_{2}}|} \text{,}
\end{equation}
where $\{ \psi_p \}$ is the set of input orbitals and $x$ is the combined space-spin variable for $(r,\sigma)$.
$W$ is the screened interaction defined as
\begin{equation}
    W_{pq,rs} = \sum_{tu} (D^{-1})_{pq,tu} v_{tu,rs} \text{,}
\end{equation}
where the dielectric function $D$ is calculated by the static response function $\chi$\cite{krauseImplementationBetheSalpeter2017,ghoshConceptsMethodsModern2016}
\begin{align}
    D_{pq,rs} &= \delta_{pr}\delta_{qs} - v_{pq,rs}\chi_{rs,rs} \text{.} \\
    \chi_{ia,ia} &= \chi_{ai,ai} =(\epsilon_i^{\text{QP}} - \epsilon_a^{\text{QP}})^{-1} \label{eq:response}
\end{align}

The BSE working equation in Eq.\ref{eq:bse} is analogous to the Casida equation\cite{casidaTimeDependentDensityFunctional1995,ullrichTimeDependentDensityFunctionalTheory2011} in TDDFT.
The only difference is that the BSE kernel replaces the XC kernel.
Thus, the scaling of solving Eq.\ref{eq:bse} is $\mathcal{O}(N^4)$ by using the canonical Davidson algorithm\cite{stratmannEfficientImplementationTimedependent1998,davidsonIterativeCalculationFew1975}. \\

\section{COMPUTATIONAL DETAILS}
We implemented the BSE/$G_{\text{RS}}W_0$ and BSE/$G_{\text{RS}}W_{\text{RS}}$ approaches in the QM4D quantum chemistry package\cite{qm4d} and applied them to calculate excitation energies of different systems.
As discussed in Ref.\citenum{liCombiningLocalizedOrbital2022},
the application of the Tamm-Dancoff approximation (TDA) improves the accuracy of BSE/$G_0W_0$ for predicting both singlet and triplet excitation energies,
because BSE/$G_0W_0$ largely underestimates the excitation energies of molecular systems.
However, for BSE/$G_{\text{RS}}W_{\text{RS}}$ and BSE/ev$GW$, which predict larger excitation energies than BSE/$G_0W_0$,
using TDA leads to similar or worse triplet results and worsens singlet results as shown Section~4 in the Supporting Information and Ref.\citenum{jacqueminBetheSalpeterFormalism2017}.
In addition,
as shown in recent studies using TDA in BSE/$GW$ can lead to blue-shifts in nanosized systems\cite{faberManybodyGreenFunction2013,roccaInitioCalculationsOptical2010,ducheminShortRangeLongRangeChargeTransfer2012} and worse accuracy for singlet-triplet energy gaps in organic molecules\cite{jacqueminBenchmarkBetheSalpeterTriplet2017}.
Therefore, TDA is not used in the present work.
We tested three different sets:
the comprehensive Truhlar-Gagliardi test set\cite{hoyerMulticonfigurationPairDensityFunctional2016} that contains singlet, triplet, valence, CT and Rydberg excitations,
the Stein CT test set\cite{steinReliablePredictionCharge2009} that contains 12 intramolecular CT excitations between an aromatic donor and the tetracyanoethylene acceptor,
and a test set for Rydberg excitations that contains three atomic systems.
For the Truhlar-Gagliardi test set\cite{hoyerMulticonfigurationPairDensityFunctional2016},
the aug-cc-pVTZ basis set\cite{dunningGaussianBasisSets1989, kendallElectronAffinitiesFirst1992} was used for all molecules, except for naphthalene, pNA and DMABN, for which the aug-cc-pVDZ basis set\cite{dunningGaussianBasisSets1989, kendallElectronAffinitiesFirst1992} was employed. It has been shown that the aug-cc-pVTZ basis sets yield converged neutral optical excitations\cite{liuAllelectronInitioBetheSalpeter2020} and even aug-cc-pVDZ results were found to deviate by not more than 0.2-0.3~eV from the basis set limit\cite{liuAllelectronInitioBetheSalpeter2020}.
B-TCNE was excluded due to the high computational cost.
Reference values for pNA and DMABN were taken from Ref.\citenum{guiAccuracyAssessmentGW2018} and from Ref.\citenum{verilQUESTDBDatabaseHighly2021} for the remaining molecules in the Truhlar-Gagliardi test set.
The reference values are the theoretical best estimates,
for example using FCI or CCSDTQ\cite{verilQUESTDBDatabaseHighly2021,guiAccuracyAssessmentGW2018}.
Geometries were all taken from Ref.\citenum{hoyerMulticonfigurationPairDensityFunctional2016}.
Note that geometries used in Ref.\citenum{verilQUESTDBDatabaseHighly2021} are slightly different from those in Truhlar-Gagliardi set. 
As shown in Ref.\citenum{loosReferenceEnergiesIntramolecular2021} and Ref.\citenum{verilQUESTDBDatabaseHighly2021}, 
the difference between theoretical best estimates obtained with two slightly different geometries is around $0.01$ \,{eV}. 
Thus, we do not expect the small differences change the conclusion.
For the Stein CT test set\cite{steinReliablePredictionCharge2009},
the cc-pVDZ\cite{dunningGaussianBasisSets1989} basis set was used.
Because theory best estimates for the Stein CT test set is not available,
the experimental values in the gas phase\cite{steinReliablePredictionCharge2009} were taken as reference,
which can be a source of errors.
For the test of Rydberg excitation energies of B$^+$, Be and Mg,
the aug-cc-pVQZ basis set\cite{dunningGaussianBasisSets1989, kendallElectronAffinitiesFirst1992} was employed.
Experimental reference values were taken from Ref.\citenum{xuTestingNoncollinearSpinFlip2014}.
TDDFT calculations were performed with the GAUSSIAN16 A.03 software\cite{g16}.
BSE/$G_0W_0$, BSE/$G_{\text{RS}}W_0$, BSE/$G_{\text{RS}}W_{\text{RS}}$ and ev$GW$ calculations were performed with QM4D.
QM4D uses Cartesian basis sets and uses the resolution of identity (RI) technique\cite{weigendAccurateCoulombfittingBasis2006,renResolutionofidentityApproachHartree2012,eichkornAuxiliaryBasisSets1995}
to compute two-electron integrals.
All basis sets and corresponding fitting basis sets were taken from the Basis Set Exchange\cite{fellerRoleDatabasesSupport1996,schuchardtBasisSetExchange2007,pritchardNewBasisSet2019}.

\section{RESULTS}

\subsection{Truhlar-Gagliardi Test Set}\label{subsec:tg_set}
\FloatBarrier

\begin{figure}
\includegraphics[width=0.9\textwidth]{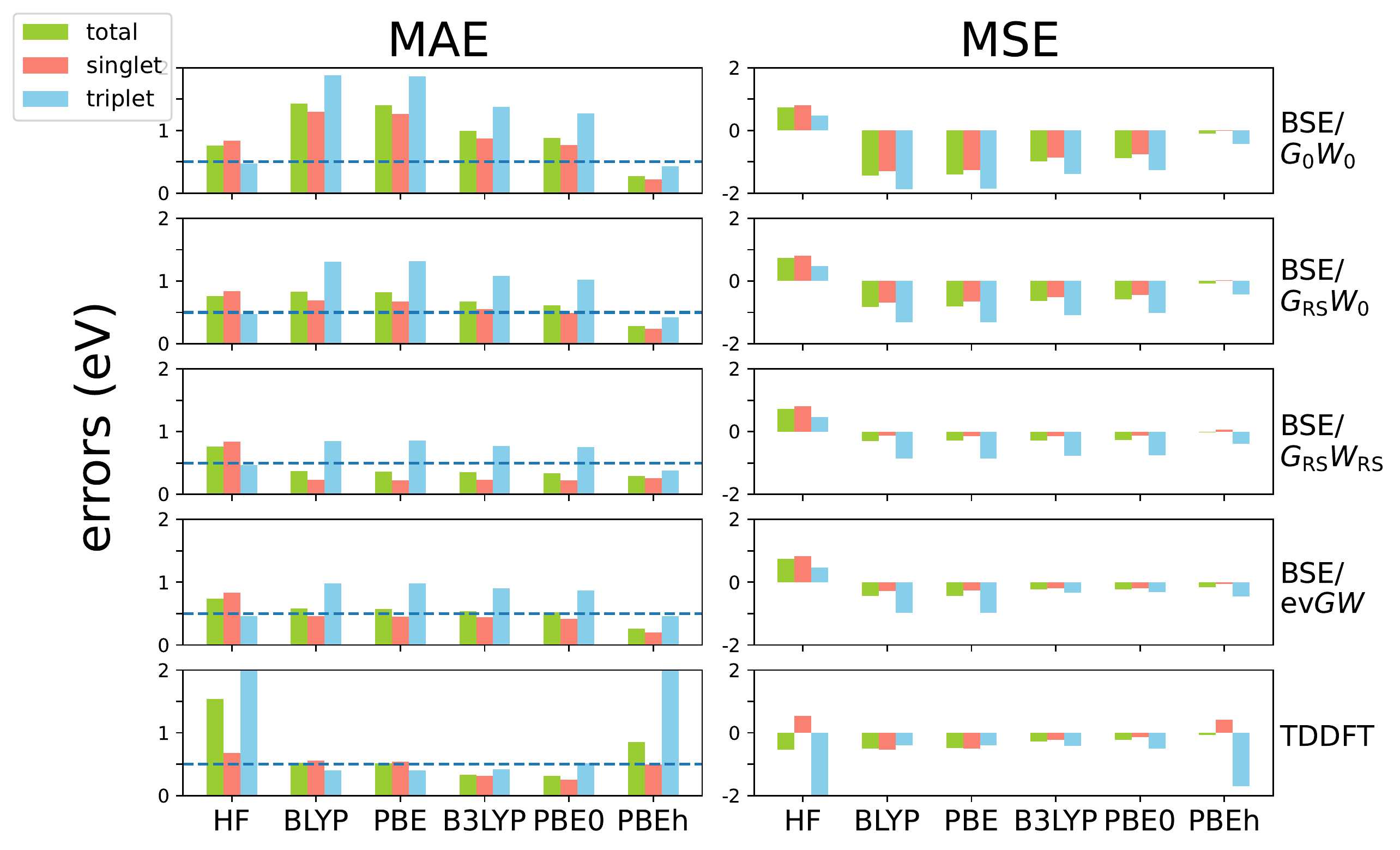}
\caption{Mean absolute errors (MAEs) and mean signed errors (MSEs) of excitation energies in Truhlar-Gagliardi test set obtained from TDDFT,
BSE/$G_0W_0$, BSE/$G_{\text{RS}}W_0$, BSE/$G_{\text{RS}}W_{\text{RS}}$ and BSE/ev$GW$ based on HF, BLYP, PBE, B3LYP, PBE0 and PBEh(0.75).
Reference values for pNA and DMABN were taken from Ref.\citenum{guiAccuracyAssessmentGW2018} and for the remaining molecules from Ref.\citenum{verilQUESTDBDatabaseHighly2021}.
The reference values are the theoretical best estimates.
The aug-cc-pVDZ basis set was used for naphthalene, pNA and DMABN, and the aug-cc-pVTZ basis set was used for the remaining systems.
B-TCNE was excluded due to the high computational cost.
Total MAEs and total MSEs were calculated by averaging all systems with equal weights. The error for system $i$ is defined as $\text{Error}_i=E^{\text{calc}}_i-E^{\text{reference}}_i$.}
\label{fig:tg}
\end{figure}

\begin{table}
    \setlength\tabcolsep{6.5pt}
    \caption{\label{tab:tg}Total mean absolute errors of excitation energies in Truhlar-Gagliardi test set obtained from TDDFT,
    BSE/$G_0W_0$, BSE/$G_{\text{RS}}W_0$, BSE/$G_{\text{RS}}W_{\text{RS}}$ and BSE/ev$GW$ based on HF, BLYP, PBE, B3LYP, PBE0 and PBEh(0.75).
    Reference values for pNA and DMABN were taken from Ref.\citenum{guiAccuracyAssessmentGW2018} and for the remaining molecules from Ref.\citenum{verilQUESTDBDatabaseHighly2021}.
    The reference values are the theoretical best estimates.
    The aug-cc-pVDZ basis set was used for naphthalene, pNA and DMABN, and the aug-cc-pVTZ basis set was used for the remaining systems.
    B-TCNE was excluded due to the high computational cost.
    Total MAEs were calculated by averaging all systems with equal weights. The error for system $i$ is defined as $\text{Error}_i=E^{\text{calc}}_i-E^{\text{reference}}_i$.}
    \begin{tabular}{c|ccccccc}
        \toprule
           & HF   & BLYP & PBE  & B3LYP & PBE0 & PBEh \\
           \midrule
    BSE/$G_0W_0$                     & 0.76 & 1.43 & 1.40 & 0.99  & 0.88 & 0.27 \\
    BSE/$G_{\text{RS}}W_0$           & 0.76 & 0.83 & 0.82 & 0.67  & 0.61 & 0.28 \\
    BSE/$G_{\text{RS}}W_{\text{RS}}$ & 0.76 & 0.37 & 0.36 & 0.35  & 0.34 & 0.29 \\
    BSE/ev$GW$                       & 0.74 & 0.48 & 0.47 & 0.52  & 0.53 & 0.26 \\
    TDDFT                            & 1.54 & 0.52 & 0.51 & 0.33  & 0.31 & 0.85 \\
    \bottomrule
    \end{tabular}
\end{table}

We first examine the performance of the BSE/$G_{\text{RS}}W_{\text{RS}}$ approach for predicting excitation energies of molecules from the Truhlar-Gagliardi test set. Excluding B-TCNE, this  test set contains 18 valence excitations,
two Rydberg excitations and two CT excitations.
The valence excitations in this set refer to $n\rightarrow \pi^*$ and $\pi \rightarrow \pi^*$ excitations.
The mean absolute errors (MAEs) and mean signed errors (MSEs) of excitation energies obtained from TDDFT, BSE/$G_0W_0$, BSE/$G_{\text{RS}}W_0$, BSE/$G_{\text{RS}}W_{\text{RS}}$ and BSE/ev$GW$ with HF, BLYP, PBE, B3LYP, PBE0 and PBEh(0.75) are shown in Fig.\ref{fig:tg} and Table.\ref{tab:tg}. The PBE-based hybrid functional PBEh(0.75) that has $75\%$ HF exchange is shown as the optimal starting point for $G_0W_0$ to predict IPs in GW100 set\cite{brunevalGWMiracleManyBody2021}.
The signed error is defined as the difference of the calculated value and the reference value, i.e., $E^{\text{calc}}-E^{\text{reference}}$.
Because this test set mainly contains valence excitations,
TDDFT with conventional DFAs provides good accuracy.
However, TDDFT based on the DFA with a large percentage of the HF exchange has large errors and can suffer from the triplet instability.
As evident from Fig.\ref{fig:tg}, BSE/$G_0W_0$ greatly underestimates the excitation energies and provides large MAEs, which was also reported previously.\cite{blaseBetheSalpeterEquation2020,liCombiningLocalizedOrbital2022}
This error can be attributed to the overscreening problem in $G_0W_0$,
i.e., the screened interaction $W$ is calculated with an underestimated KS gap\cite{blaseBetheSalpeterEquation2020}.
The underestimated fundamental gap in $G_0W_0$ leads to the underestimated optical gap in BSE/$G_0W_0$.
In addition,
BSE/$G_0W_0$ has a strong starting point dependence.
The difference between MAEs of BSE/$G_0W_0$ with GGA and hybrid functionals is larger than $0.7$ \,{eV}.
BSE/$G_0W_0$ based on PBEh(0.75) that is the optimal starting point for valence QP energy provides a small MAE of $0.27$ \,{eV}.
However,
the percentage of the HF exchange needs to be re-optimized for excitations of different characters and different species\cite{yaoAllElectronBSEGW2022}.
Recent work has shown that the optimally tuned range-separated hybrid DFAs can be a good starting point for BSE/$G_0W_0$\cite{mckeonOptimallyTunedRangeseparated2022}.
The BSE/$G_{\text{RS}}W_0$ approach improves upon BSE/$G_0W_0$.
Using $G_{\text{RS}}W_0$ instead of $G_0W_0$, the MAEs are reduced by around $0.6$ \,{eV} with GGA functionals and by around $0.3$ \,{eV}  with hybrid functionals.
However, there is still an undesired starting point dependence in BSE/$G_{\text{RS}}W_0$,
because the screened interaction in $G_{\text{RS}}W_0$ is calculated at the chosen DFA level.
The BSE/$G_{\text{RS}}W_{\text{RS}}$ approach significantly outperforms BSE/$G_0W_0$ and BSE/$G_{\text{RS}}W_0$.
The MAEs from BSE/$G_{\text{RS}}W_{\text{RS}}$ with conventional DFAs are around $0.4$ \,{eV}.
They are similar to the MAEs from TDDFT with hybrid functionals and are slightly better than the ones from BSE/ev$GW$.
BSE/$G_{\text{RS}}W_{\text{RS}}$ with the optimal starting point PBEh(0.75) provides the smallest MAE of $0.29$ \,{eV}.
As shown in Section~3 in the Supporting Information,
fundamental gaps obtained from RS orbital energies are always larger than those obtained from KS orbital energies.
By inserting the RS Green's function into the RPA equation to formulate the screened interaction,
BSE/$G_{\text{RS}}W_{\text{RS}}$ greatly reduces the overscreening error and provides excellent accuracy.
The starting point dependence in BSE/$G_{\text{RS}}W_{\text{RS}}$ is largely reduced,
which is similar to BSE/ev$GW$.
The different DFAs induce only small changes of less than 0.1~eV in the BSE/$G_{\text{RS}}W_{\text{RS}}$ MAEs.
We find that BSE/$G_{\text{RS}}W_{\text{RS}}$, BSE/$G_0W_0$ and BSE/ev$GW$ yield triplet excitation energies which are significantly too low,
which is in agreement with previous work\cite{blaseBetheSalpeterEquation2020}.

\FloatBarrier

\subsection{Stein CT Test Set}\label{subsec:stein_set}

\FloatBarrier

\begin{table*}
\setlength\tabcolsep{2.5pt}
    \caption{\label{tab:stein}Mean absolute errors (MAEs) and mean signed errors (MSEs) of CT excitation energies in Stein CT test set obtained
    from TDDFT, BSE/$G_0W_0$, BSE/$G_{\text{RS}}W_0$ and BSE/$G_{\text{RS}}W_{\text{RS}}$ with HF, BLYP, PBE, B3LYP, PBE0 and PBEh(0.75).
    All values in eV.
    Geometries were taken from Ref.\citenum{steinReliablePredictionCharge2009}.
    Experiment values in the gas phase were taken as the reference values\cite{steinReliablePredictionCharge2009}.
    Gas phase experimental references were used. The error for system $i$ is defined as $\text{Error}_i=E_i^{\text{calc}}-E_i^{\text{experiment}}$
    The cc-pVDZ basis set was used.}
\begin{tabular*}{0.99\linewidth}{ccccccccccccc}
\toprule
            & \multicolumn{2}{c}{HF} & \multicolumn{2}{c}{BLYP} & \multicolumn{2}{c}{PBE} & \multicolumn{2}{c}{B3LYP} & \multicolumn{2}{c}{PBE0} & \multicolumn{2}{c}{PBEh}\\
            \cmidrule(l{0.5em}r{0.5em}){2-3}\cmidrule(l{0.5em}r{0.5em}){4-5} \cmidrule(l{0.5em}r{0.5em}){6-7} \cmidrule(l{0.5em}r{0.5em}){8-9} \cmidrule(l{0.5em}r{0.5em}){10-11} \cmidrule(l{0.5em}r{0.5em}){12-13}
            & MAE & MSE & MAE & MSE & MAE & MSE & MAE & MSE & MAE & MSE & MAE & MSE \\ \hline
TDDFT                            & 0.78 & 0.78  & 1.44 & -1.44 & 1.45 & -1.45 & 1.16 & -1.16 & 1.08 & -1.08 & 0.19 & 0.02     \\
BSE/$G_0W_0$                     & 0.10 & -0.06 & 1.28 & -1.28 & 1.31 & -1.31 & 0.74 & -0.74 & 0.65 & -0.65 & 0.14 & 0.09     \\
BSE/$G_{\text{RS}}W_0$           & 0.10 & -0.06 & 0.35 & -0.35 & 0.35 & -0.37 & 0.29 & -0.28 & 0.29 & -0.29 & 0.16 & -0.15    \\
BSE/$G_{\text{RS}}W_{\text{RS}}$ & 0.10 & -0.06 & 0.17 & 0.14  & 0.18 & 0.11  & 0.14 & 0.09  & 0.17 & 0.14  & 0.11 & -0.07    \\
\bottomrule
\end{tabular*}
\end{table*}

We further study the performance of BSE/$G_{\text{RS}}W_{\text{RS}}$ on predicting CT excitation energies by testing Stein CT test set.
This test set contains 12 intramolecular CT systems.
The MAEs and MSEs of excitation energies obtained from TDDFT, BSE/$G_0W_0$, BSE/$G_{\text{RS}}W_0$ and BSE/$G_{\text{RS}}W_{\text{RS}}$ with HF, BLYP, PBE B3LYP, PBE0 and PBEh(0.75) are listed in Table~\ref{tab:stein}.
It shows that TDDFT with conventional DFAs fails to predict CT excitation energies due to the incorrect description of the long-range behavior.
TDDFT with both GGA and hybrid functionals greatly underestimates the CT excitation energies and gives MAEs larger than $1.0$ \,{eV}.
TDDFT with PBEh(0.75) that has a large percentage of the HF exchange provides a small MAE of $0.19$\,{eV}.
In addition, TDDFT has a strong starting point dependence.
BSE/$G_0W_0$ provides improved results over TDDFT because of the correct long-range behavior from the screened interaction.
However,
BSE/$G_0W_0$ still suffers from a strong dependence on the choice of the DFA and yields relatively large errors.
As shown in Section~6 in the Supporting Information,
BSE/$G_0W_0$ always underestimates the excitation energies of CT systems. Slightly larger excitation energies can be obtained when using the TDA.\cite{liCombiningLocalizedOrbital2022}
The BSE/$G_{\text{RS}}W_0$ approach improves again upon BSE/$G_0W_0$.
Compared with BSE/$G_0W_0$,
the MAEs of BSE/$G_{\text{RS}}W_0$ are  $1.0$ \,{eV} and $0.4$ \,{eV} smaller with GGA and hybrid functionals, respectively.
The dependence on the DFA is reduced to only $0.09$ \,{eV} in the BSE/$G_{\text{RS}}W_0$ scheme.
The BSE/$G_{\text{RS}}W_{\text{RS}}$ approach provides the most accurate results with the smallest starting point dependence.
The MAEs of BSE/$G_{\text{RS}}W_{\text{RS}}$ with all tested DFAs are only around $0.15$ \,{eV},
which are comparable to the accuracy of BSE/ev$GW$ as reported in Ref.\citenum{blaseChargetransferExcitationsMolecular2011}.
In addition,
the dependence on the choice of the DFA in BSE/$G_{\text{RS}}W_{\text{RS}}$ is largely eliminated.
The difference originating from using different DFAs is only around $0.04$ \,{eV}.

\FloatBarrier

\subsection{Rydberg Excitations}\label{subsec:rydberg}

\FloatBarrier

\begin{figure}
\includegraphics[width=0.9\textwidth]{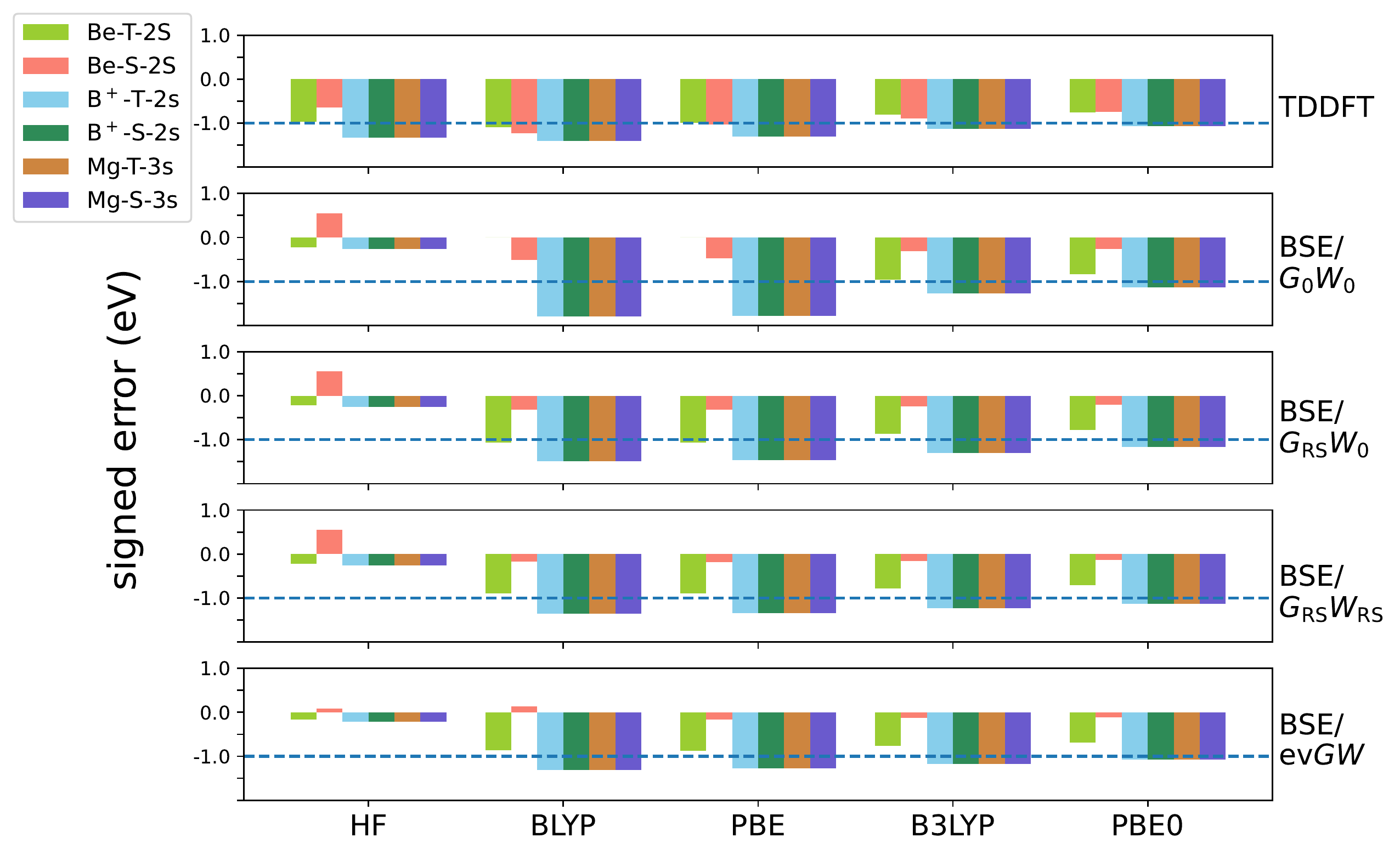}
\caption{Signed errors of B$^+$, Be and Mg obtained from TDDFT, BSE/$G_0W_0$, BSE/$G_{\text{RS}}W_0$, BSE/$G_{\text{RS}}W_{\text{RS}}$ and BSE/ev$GW$ with HF, BLYP, PBE, B3LYP and PBE0.
All values in eV.
Experiment vales were taken as the reference values\cite{xuTestingNoncollinearSpinFlip2014}.
The aug-cc-pVQZ basis set was used.
The signed error for system $i$ is defined as $\text{Error}_i=E^{\text{calc}}_i-E^{\text{experiment}}_i$.}
\label{fig:rydberg}
\end{figure}

We further investigate the performance of the BSE/$G_{\text{RS}}W_0$ and the BSE/$G_{\text{RS}}W_{\text{RS}}$ approaches for predicting Rydberg excitation energies of atomic systems by testing B$^+$, Be and Mg.
The signed errors of Rydberg excitation energies obtained from TDDFT, BSE/$G_0W_0$, BSE/$G_{\text{RS}}W_0$, BSE/$G_{\text{RS}}W_{\text{RS}}$ and BSE/ev$GW$ with HF, BLYP, PBE, B3LYP and PBE0 are listed in Figure~\ref{fig:rydberg}.
Numerical results are shown in the Section~7 in the Supporting Information.
Similar to the CT excitation energies,
it is well-known that TDDFT with common DFAs has relatively large errors for predict Rydberg excitation energies\cite{blaseBetheSalpeterEquation2020}. The latter are largely underestimates with GGA as well as hybrid functionals in TDDFT.
Compared to TDDFT, the BSE/$G_0W_0$ approach yields only slightly better predictions, reducing the MAE  by around $0.1$ \,{eV}.
TDDFT and BSE/$G_0W_0$ show both a strong starting point dependence.
BSE/$G_{\text{RS}}W_0$ slightly improves upon BSE/$G_0W_0$, reducing the MAE by $0.1$ \,{eV} with respect to BSE/$G_0W_0$. The BSE/$G_{\text{RS}}W_{\text{RS}}$ approach provides further improvements over BSE/$G_{\text{RS}}W_0$.
As shown in Section~7 in the Supporting Information and Fig~\ref{fig:rydberg},
BSE/$G_{\text{RS}}W_{\text{RS}}$ shows significant improvements over TDDFT for predicting singlet Rydberg excitation energies with reduced errors around $0.4$ \,{eV}.
However,
the improvements on triplet Rydberg excitation energies are small.
The MAEs of BSE/$G_{\text{RS}}W_{\text{RS}}$ with all different DFAs are around $0.6$ \,{eV},
which is close to the BSE/ev$GW$ level. The DFA starting point dependence is reduced to round $0.2$ \,{eV} and  $0.1$ \,{eV} with BSE/$G_{\text{RS}}W_0$ and BSE/$G_{\text{RS}}W_{\text{RS}}$, respectively.

\FloatBarrier

\section{Conclusion}\label{sec:conclusion}
In this work,
we applied the BSE formalism on top of the  $G_{\text{RS}}W_{\text{RS}}$ method to calculate valence, Rydberg and charge transfer excitation energies of molecular systems.
The $G_{\text{RS}}W_{\text{RS}}$ method
provides improved fundamental gaps compared to $G_0W_0$ and largely reduces the dependence on the choice of the density functional approximation.
In the BSE/$G_{\text{RS}}W_{\text{RS}}$ approach,
the quasiparticle energies from $G_{\text{RS}}W_{\text{RS}}$ are used in BSE.
For the Truhlar-Gagliardi test set,
we found that BSE/$G_{\text{RS}}W_{\text{RS}}$ provides excellent accuracy for excitations of different characters (valence,CT and Rydberg excitations) with MAEs around $0.4$ \,{eV}.
The accuracy of BSE/$G_{\text{RS}}W_{\text{RS}}$ is similar to time-dependent density functional theory (TDDFT) and slightly better than BSE/ev$GW$.
Using the Stein CT test set,
we further showed that  BSE/$G_{\text{RS}}W_{\text{RS}}$ is significantly more accurate for predicting CT excitation energies than  BSE/$G_0W_0$ and TDDFT. We also found that the predictions compare well to BSE/ev$GW$ results reported in the literature\cite{blaseChargetransferExcitationsMolecular2011}.
We also showed that BSE/$G_{\text{RS}}W_{\text{RS}}$ predict accurate Rydberg excitation energies for atomic systems. We found for all three test sets that the dependence on the choice of the DFA is also largely eliminated.
The computational cost of BSE/$G_{\text{RS}}W_{\text{RS}}$ is similar to BSE/$G_0W_0$,
which has a much lower computational cost than BSE/ev$GW$. \\

This work demonstrates the BSE/$G_{\text{RS}}W_{\text{RS}}$ approach is accurate and efficient for predicting all three types of excitation energies of a broad range of systems.
Therefore, the BSE/$G_{\text{RS}}W_{\text{RS}}$ approach is expected to expand the applicability of the BSE/$GW$ approach. \\

\begin{acknowledgement}
J. L acknowledges the support from the National
Institute of General Medical Sciences of the National Institutes of
Health under award number R01-GM061870. D.G. acknowledges the Emmy Noether Programme of the German Research Foundation under
project number 453275048. W.Y. acknowledges the support
from the National Science Foundation (grant no. CHE-1900338).
\end{acknowledgement}

\section*{SUPPORTING INFORMATION}
See the Supporting Information for fundamental gaps obtained from different $GW$ method, errors of using the linearized quasiparticle equation in BSE, fundamental gaps obtained from KS-DFT and KS-DFT with RS, comparison of excitation energies obtained from BSE/$G_{\text{RS}}W_{\text{RS}}$ with and without the TDA, results of Truhlar-Gagliardi test set, results of Stein CT test set, results of Rydberg excitation energies.

\section*{Data Availability Statement}
The data that support the findings of this study are available from the corresponding author
upon reasonable request.

\bibliography{ref, software}

\end{document}